\documentclass
[
amssymb,prl,aps,twocolumn,floatfix,tightenlines,superscriptaddress]{revtex4}
\pdfoutput=1
\usepackage{graphicx}
\usepackage{color}
\usepackage{eurosym}
\usepackage{dcolumn}
\usepackage{bm}
\usepackage{subfigure}
\usepackage[final]{pdfpages}
\usepackage{setspace}
\usepackage{pdfpages}
\graphicspath{{./images/}}

\newcommand{\ket}[1]{\left | #1 \right \rangle}
\newcommand{\bra}[1]{\left \langle #1 \right |}

\begin{document}

\title{Observing fermionic statistics with photons in arbitrary processes}

\author{Jonathan C. F. Matthews}
\affiliation{Centre for Quantum Photonics, H. H. Wills Physics Laboratory \& Department of Electrical and Electronic Engineering, University of Bristol, Merchant Venturers Building, Woodland Road, Bristol, BS8 1UB, UK}
\altaffiliation{These authors contributed equally}
\author{Konstantinos Poulios}
\affiliation{Centre for Quantum Photonics, H. H. Wills Physics Laboratory \& Department of Electrical and Electronic Engineering, University of Bristol, Merchant Venturers Building, Woodland Road, Bristol, BS8 1UB, UK}
\altaffiliation{These authors contributed equally}
\author{Jasmin D. A. Meinecke}
\affiliation{Centre for Quantum Photonics, H. H. Wills Physics Laboratory \& Department of Electrical and Electronic Engineering, University of Bristol, Merchant Venturers Building, Woodland Road, Bristol, BS8 1UB, UK}
\altaffiliation{These authors contributed equally}
\author{Alberto Politi}
\affiliation{Centre for Quantum Photonics, H. H. Wills Physics Laboratory \& Department of Electrical and Electronic Engineering, University of Bristol, Merchant Venturers Building, Woodland Road, Bristol, BS8 1UB, UK}
\affiliation{Presently at Center for Spintronics and Quantum Computation, University of California Santa Barbara, USA}
\altaffiliation{These authors contributed equally}
\author{Alberto Peruzzo}
\affiliation{Centre for Quantum Photonics, H. H. Wills Physics Laboratory \& Department of Electrical and Electronic Engineering, University of Bristol, Merchant Venturers Building, Woodland Road, Bristol, BS8 1UB, UK}
\author{Nur Ismail}
\author{Kerstin W\"{o}rhoff}
\affiliation{\mbox{Integrated Optical Microsystems Group, MESA+ Institute for Nanotechnology, University of Twente, Enschede, The Netherlands}}
\author{Mark G. Thompson}
\author{Jeremy L. O'Brien}
\email{jeremy.obrien@bristol.ac.uk}
\affiliation{Centre for Quantum Photonics, H. H. Wills Physics Laboratory \& Department of Electrical and Electronic Engineering, University of Bristol, Merchant Venturers Building, Woodland Road, Bristol, BS8 1UB, UK}

\begin{abstract}
Quantum mechanics defines two classes of particles---bosons and fermions---whose exchange statistics fundamentally dictate the dynamics of quantum systems. Here we develop a scheme that uses multi-partite entanglement to directly observe the correlated detection statistics of any number of fermions in any physical process. This approach relies on sending each of the entangled particles through identical copies of the process and by controlling a single phase parameter in the entangled state, the correlated detection statistics can be continuously tuned between bosonic and fermionic statistics. We implement this scheme via two entangled photons shared across the  polarisation modes of a single photonic chip to directly mimic the fermion, boson and intermediate behaviour of two-particles undergoing a continuous time quantum walk with an average similarity of 93.6$\pm$0.2\% with ideal models. The ability to simulate fermions with photons is likely to have applications for verifying computationally hard boson scattering in linear optics and for observing particle correlations (including fractional statistics) in analogue simulation using any physical platform that can prepare the entangled state prescribed here.
\end{abstract}

\maketitle

\setstretch{1}

A particularly striking departure from classical physics in quantum mechanics is the existence of two fundamental particle classes characterised by their behaviour under pairwise exchange ($a^\dagger_j a^\dagger_k =\pm a^\dagger_k a^\dagger_j$). For example an arbitrary number of bosons (obeying Bose-Einstein statistics: `$+$') can occupy a single quantum state, leading to bunching \cite{ho-prl-59-2044} and Bose-Einstein condensation; whereas fermions (obeying Fermi-Dirac statistics: `$-$') are strictly forbidden to occupy the same state via the Pauli-exclusion principle \cite{li-nat-391-263}. Behaviour intermediate between bosons and fermions can also be considered; particles confined to two dimensions known as anyons \cite{wi-prl-48-1144} are predicted to exhibit intermediate behaviour with fractional exchange statistics ($a^\dagger_j a^\dagger_k = e^{i\phi} a^\dagger_k a^\dagger_j$, $0\!<\!\phi\!<\!\pi$)---a concept applied in superconductivity \cite{Wilczek} and used to explain the fractional quantum Hall effect \cite{la-prl-50-1395}.

\begin{figure}[tp!]
    \centering
    \includegraphics[width = 6cm]{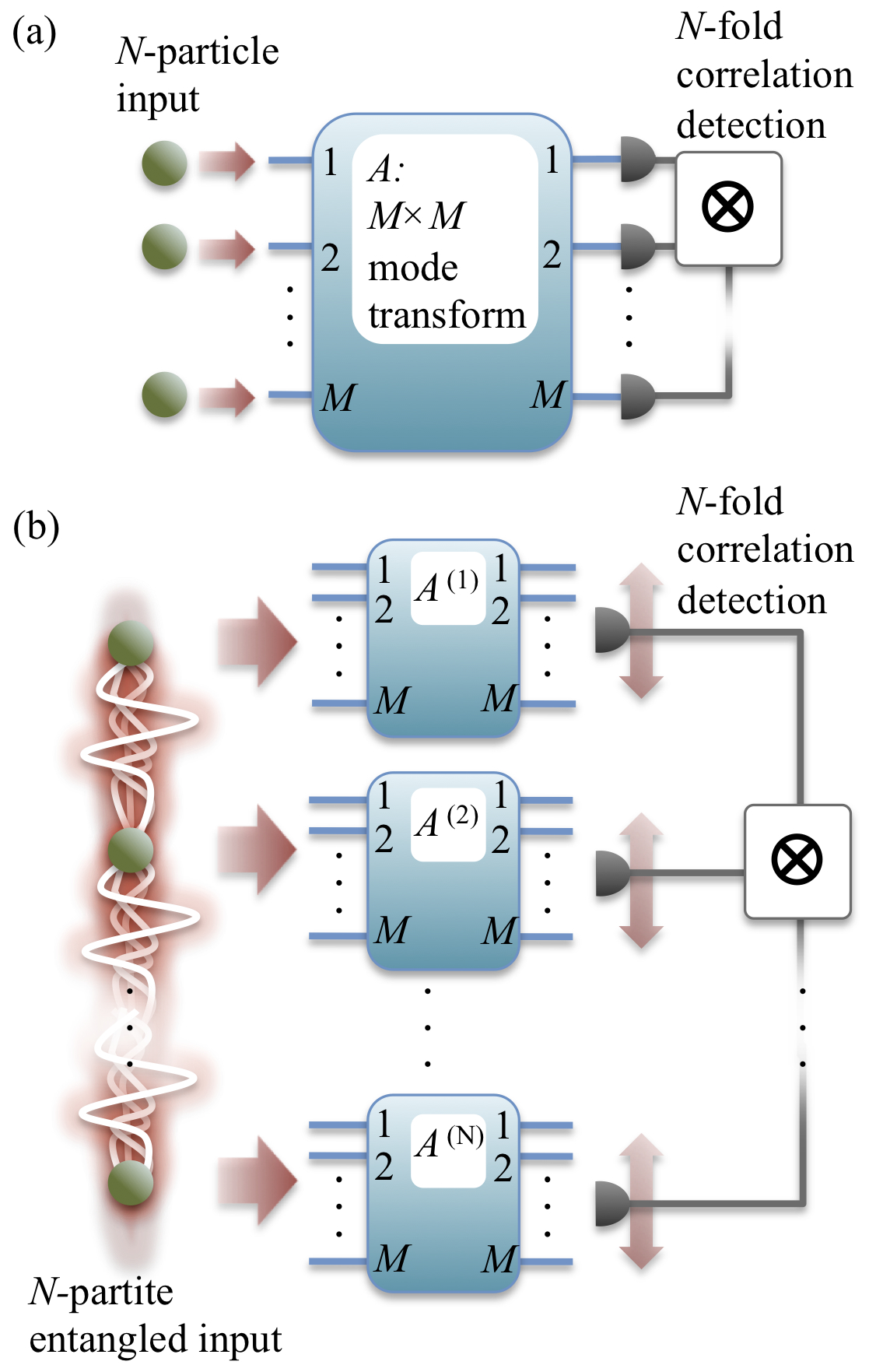}
\caption[]{\footnotesize{Simulating quantum interference of quantum particles. (a) Quantum interference of $N$ identical particles with arbitrary statistics launched into a network described by matrix $A$ leads to quantum correlated detection events at the output. (b) Experimental simulation of quantum interference can be directly achieved by sharing $N$-partite, $N$-level entanglement across $N$ copies of $A$, where the choice of simulated particle statistics is controlled by the phase of initial entanglement.}}
\label{fig_EntangPic}
\end{figure}

Here we show that by tuning one phase parameter, entanglement and a single experimental setup can exactly simulate the measurement statistics of an arbitrary mode transformation $A$ (Fig.~\ref{fig_EntangPic} (a)) of $N$ identical fermions, bosons or particles with fractional statistics. 
It is well known that the symmetric and anti-symmetric states $\ket{\psi^{\pm}} = \ket{H}_1\ket{V}_2 \pm \ket{V}_1\ket{H}_2$ of two polarised photons input into a $50\%$ reflectivity beam splitter exhibits fermion-like anti-bunching and boson-like bunching, respectively \cite{ze-ps-T76-203}. Furthermore, intermediate behaviour for a beamsplitter has been observed by controlling a phase to tune between the two Bell states $\ket{\psi^+}$ and $\ket{\psi^{-}}$ \cite{mi-pra-53-R1209,wa-epjd-42-179}. It is less widely appreciated that the resulting detection statistics can also be exactly mimicked with two photon path-number entangled NOON states \cite{ra-prl-65-1348}; however this behaviour does not generalise to processes other than beamsplitters, the simplest example being the identity operation for which bunching always occurs regardless of the phase of the NOON state.
The approach we present shares $N$-partite, $N$-level entanglement, parameterised by phase $\phi$, across $N$ copies of $A$ (Fig.~\ref{fig_EntangPic} (b)).

After presenting the scheme in the case of two particles in an arbitrary linear evolution, we experimentally demonstrate the simulation of arbitrary particle statistics with a complex photonic network---a continuous time quantum walk \cite{pe-prl-100-170506,pe-sci-329-1500}.We achieve the required entangled state with post-selected polarisation entanglement shared across the TE and TM modes of an evanescently coupled waveguide array. This enables us to directly observe the evolution of symmetric and anti-symmetric two particle wave functions in a coupled oscillator Hamiltonian by switching the phase of the entangled state between $0$ and $\pi$. Mixed-symmetric quantum walk behaviour is also observed by tuning the entangled state's phase between $0$ and $\pi$, corresponding to behaviour intermediate between bosons and fermions. We note the fermion simulation directly maps to the dynamics of two excitations in a spin$-\frac{1}{2}$ chain consisting of ten sites: Using the Jordan-Wigner transformation, excitations correspond to fermion creation operators and the vacuum to the ground state of each spin \cite{hi-pra-75-062321}. We finally provide a mathematical demonstration of our scheme in the case of an arbitrary number of particles evolving through any linear Hamiltonian. It is important to note that neither the two-particle case, nor the N-particle generalization, require any post-selection and are, in princlple, scalable.

\section*{Results}

Exchange statistics refer to a wave-function acquiring phase $\phi$ after interchanging two indistinguishable quantum particles. Occupation of labeled modes $j$ are modelled by annihilation operators $a_j$ and creation operators  $a^\dagger_j$ that obey the commutation relations
\begin{eqnarray}
&&a^\dagger_j a^\dagger_k - e^{i\phi} a^\dagger_k a^\dagger_j = 0 \label{com1mt}\\ 
&&a_j a_k - e^{i\phi} a_k a_j = 0\label{com2mt}\\
&&a_j a^\dagger_k - e^{i\phi} a^\dagger_k a_j = \delta_{j,k} \label{com3mt}
\end{eqnarray}
for $j<k$. In three dimensions particles are either bosons ($\phi = 0$) or fermions ($\phi = \pi$) and successive swapping is represented by the permutation group while the correlated detection statistics are associated respectively to the permanent and determinant of a given matrix \cite{li-njp-7-155}. In one- and two-dimensions anyons \cite{wi-prl-48-1144} can exist where swapping two identical anyons $\omega$ times (represented by the braid group) results in the combined wave-function acquiring phase $\phi = \omega \theta$ where $0<\theta<\pi$ and the positive (or negative) winding number $\omega$ keeps track of the number of times two anyons braid in trajectory space with clockwise (or anti-clockwise) orientation \cite{Wilczek}. 

The quantum correlation function for two identical non-interacting particles (input into $j$ and $k$ of $A$, and detected at outputs $r$ and $q$) is given by
\begin{eqnarray}
\Gamma_{r,q}^\phi = \left| A_{r,j} A_{q,k} + e^{i\phi} A_{r,k} A_{q,j} \right|^2,
\label{2_part_correlation_fn}
\end{eqnarray}
for bosons \cite{br-prl-102-253904} ($\phi = 0$) and fermions \cite{li-njp-7-155} ($\phi = \pi$). Here, the process $A$ is represented by a matrix whose element $A_{r,j}$ is the complex transition amplitude from $j$ to $r$. Non-classical behaviour is evident in the constructive and destructive interference between typically complex terms inside the $\left| \cdot \right|^2$, dictated by $\phi$ $=\omega \theta$. By defining  fractional exchange algebra (\ref{com1mt}-\ref{com3mt}) for $0< \theta$ $<\pi$ and some integer winding number $\omega$, we can observe the effect of fractional exchange statistics on quantum interference between two indistinguishable anyons undergoing braiding operations dependent upon $A$; 
Eq.~(\ref{2_part_correlation_fn}) represents the resulting quantum interference between two abelian anyons ($0< \theta$ $<\pi$) in superposition of braiding $\omega$ times and not braiding at all. To see this we define without loss of generality the input labels by $j<k$ and the output labels by $r<q$ and consider two identical anyons created at modes $j$ and $k$,  transformed according to $A$ and detected at modes $r$ and $q$. Correlated detection at $r$ and $q$ reveals two indistinguishable possibilities that correspond to the two terms in Eq.~(\ref{2_part_correlation_fn}): $A$ maps the anyon at $j$ to mode $r$ and the anyon at $k$ to mode $q$ with amplitude $A_{r,j} A_{q,k}$; or $A$ maps the anyon at $j$ to mode $q$ and the anyon at $k$ to mode $r$ via $\omega$ clockwise braids with fractional phase $\phi$ and amplitude $A_{q,j} A_{r,k}$. From this definition, we use relations (\ref{com1mt}-\ref{com3mt}) to derive expression (\ref{2_part_correlation_fn}) for $0<\theta<\pi$ and $\phi = \omega \theta$ (see Supplementary Information). 

\subsection*{Simulation of two-particle statistics}

To physically simulate quantum interference given by Eq.~(\ref{2_part_correlation_fn}), we use two copies of $A$, denoted $A^{a}$ and $A^{b}$, together with bi-partite entanglement of the form $\ket{\psi\!\left(\phi\right)} = ({a}^\dagger_j {b}^\dagger_k + e^{i \phi}{a}^\dagger_k {b}^\dagger_j)\ket{0}/\sqrt{2} $ where creation operators ${a_j}^\dagger$ and $b_k^\dagger$ are associated to identical networks $A^{a}$ and $A^{b}$
respectively; creation operator sub-indices denote the modes on which the creation operators act. Once $\ket{\psi\!\left(\phi\right)}$ is generated, the class of input entangled particles becomes irrelevant; only one particle is present in each of $A^{a}$ and $A^b$, therefore no further quantum interference between particles will occur. This implies application with other quantum platforms, using for example entangled fermions to simulate Bose-Einstein quantum interference. Evolution of the two particle state through the tensor product of transformations is therefore modelled by
\begin{eqnarray}
\ket{\psi\!\left(\phi\right)} \stackrel{A^{a} \otimes A^{b}}{\longrightarrow}\sum_{s,t}\left(\frac{A^{a}_{s,j}A^{b}_{t,k} \!+\! e^{i \phi} A^{a}_{s,k} A^{b}_{t,j}}{\sqrt{2}}\right) {a^\dagger_{s}} {b^\dagger_{t}}\ket{0}
\label{emulation_formula}
\end{eqnarray} 
It follows that correlated detection probability across outputs $r$ of $A^{a}$ and $q$ of $A^{b}$ simulates Eq.~(\ref{2_part_correlation_fn}) according to
\begin{eqnarray}
P^{\phi}_{r,q}&=&\frac{1}{2}\big|\bra{0} {a_r}{b_q}\sum_{s,t} \left(A^{a}_{s,j} A^{b}_{t,k} +e^{i\phi} A^{a}_{s,k} A^{b}_{t,j}\right) {a_s}^\dagger{b_t}^\dagger  \ket{0}\big|^2
\nonumber \\
&=& \frac{1}{2}\left|A^{a}_{r,j} A^{b}_{q,k} + e^{i \phi} A^{a}_{r,k} A^{b}_{q,j}\right|^2 \equiv \frac{1}{2}\Gamma^{\phi}_{r,q}
\label{emulationeqn}
\end{eqnarray}
where the $1/2$ pre-factor is due to normalisation of the initial entanglement. 
Since the phase of the entangled state can be defined as $\phi = \omega \theta$, this method simulates the interference statistics of two anyons of fractional phase $\theta$ that have not braided, with two anyons also characterised by $\theta$ that have braided $\omega$ times (equivalently anyons that are in a superposition of braiding $\omega_1$ and $\omega_2$ times for $\phi=(\omega_1-\omega_2)\theta$---the ideal statistics are the same).

\subsection*{Experiment}

\begin{figure}[tp!]
    \centering
    \includegraphics[width = 8cm]{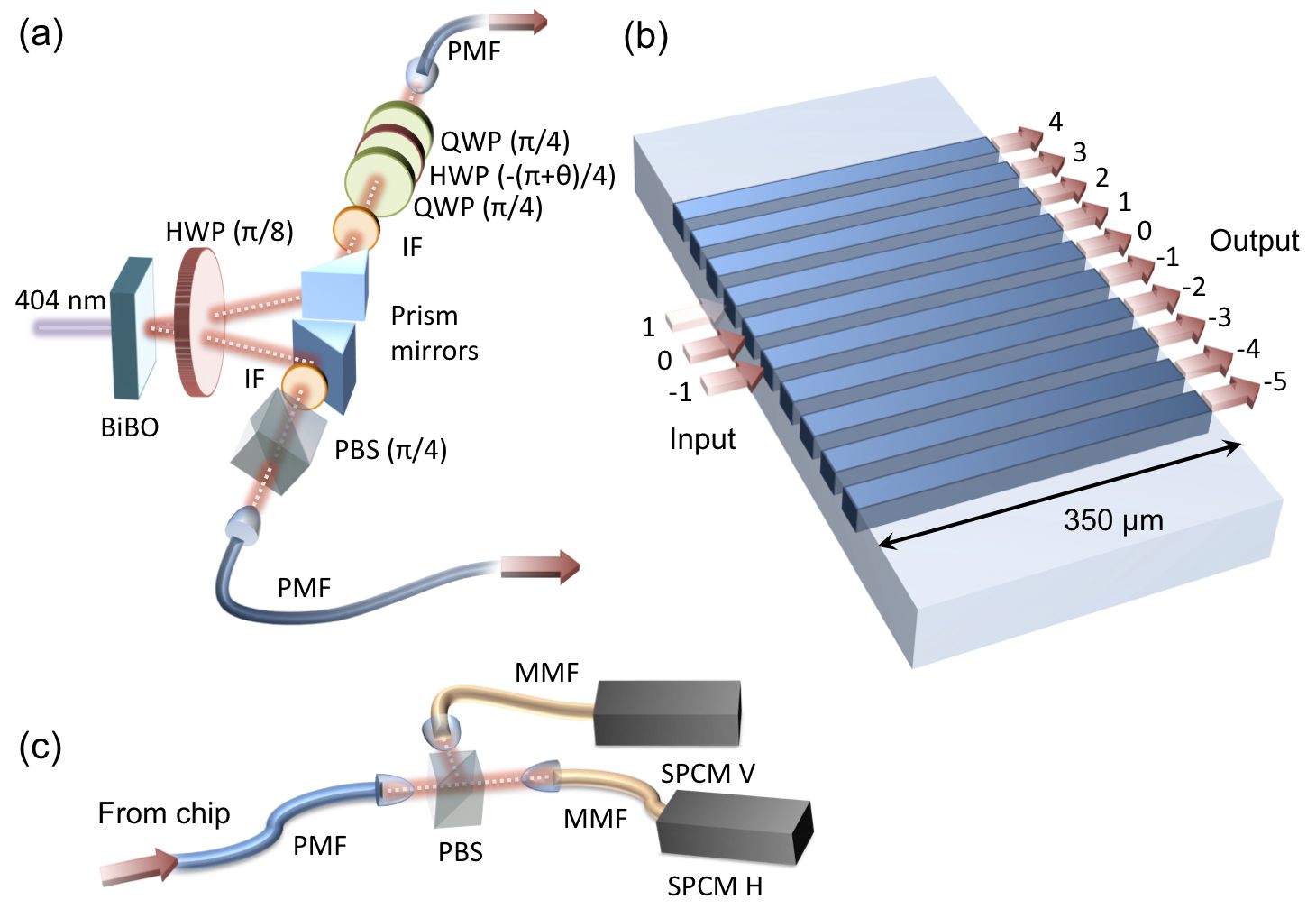}
\caption[]{\footnotesize{{Experimental setup.} (a) The parametric down-conversion based photon pair source with quarter-waveplates (QWP), half-waveplates (HWP) and a polarisation beamsplitter cube (PBS) used as a polariser, to control the initial state before input into the chip. Down conversion in the nonlinear bismuth borate (BiBO) crystal yields degenerate photon pairs (808nm, filtered with 2nm full width, half maximum interference filters, IF) collected into polarisation maintaining fibres (PMF). See Supplementary Information for further details. (b)  Two copies of the quantum walk unitary are realised by accessing the TE and TM modes of the polarisation preserving waveguide array. Three input modes of the array are used in the experiment (-1,0,1) accessed via waveguide bends that fan to a pitch of 250$\mu$m and butt-coupled to an array of PMF. Output is accessed via a fan of waveguide from the coupling region to a pitch of 125$\mu$m. See Supplementary Information for details. (c) The detection scheme consists of five PBS and single photon counting module (SPCM) pairs. PMF coupled to the chip launch the output photons onto each PBS which separates TE and TM modes guided in the waveguide. The two outputs of each PBS are collected into multi-mode fibre (MMF) coupled to SPCMs for correlated detection via counting electronics.
}}
\label{fig_ExpSetup}
\end{figure}

We experimentally simulated Eq.~(\ref{2_part_correlation_fn}) for arbitrary $\phi$ in a coupled oscillator Hamiltonian equivalent to sharing entanglement across two continuous time quantum walks, realised with arrayed waveguides \cite{pe-prl-100-170506} (see Supplementary Information). We measured the correlations of the polarisation entangled state  $(\ket{H}_j\ket{V}_k + e^{i \phi}\ket{V}_j\ket{H}_k)/\sqrt{2}$ shared across two copies of a process $A$: $A^a$ and $A^b$, realised as the TE and TM modes of a waveguide array fabricated in silicon oxynitride (SiO$_x$N$_y$) (Fig.~\ref{fig_ExpSetup} (b), see also Supplementary Information). The waveguides and butt-coupled polarisation maintaining fibres (PMFs) are birefringent to ensure negligible cross-talk between horizontal and vertical polarisations. Detection statistics of the output photons in the $\{\ket{H},\ket{V}\}$ basis were analysed using polarising beam splitters (PBS) and pairs of multi-mode fibre coupled avalanche photo-diode single photon counting modules (SPCM) (Fig.~\ref{fig_ExpSetup} (c)). 
By inputting the state $(\ket{H}_j + \ket{V}_j)(\ket{H}_k + e^{-i \phi} \ket{V}_k)/2$ and detecting one $\ket{H}$ photon and one $\ket{V}$ photon at any pair of outputs of $A^a$ and $A^b$ we post-selected the input state $(\ket{H}_j\ket{V}_k + e^{i (\phi + \epsilon)}\ket{V}_j\ket{H}_k)/{\sqrt{2}}$, with a measurable phase off-set $\epsilon$ (from birefringence of PMF and the waveguide circuit). This phase $\epsilon$ was compensated using a polarisation phase rotation (Fig.~\ref{fig_ExpSetup} (a)). Note that because of this compensation, the approach described is immune to birefringence of the 2m PMF and input waveguides.

\begin{figure}[htp!]
    \centering
    \includegraphics[width = 8cm]{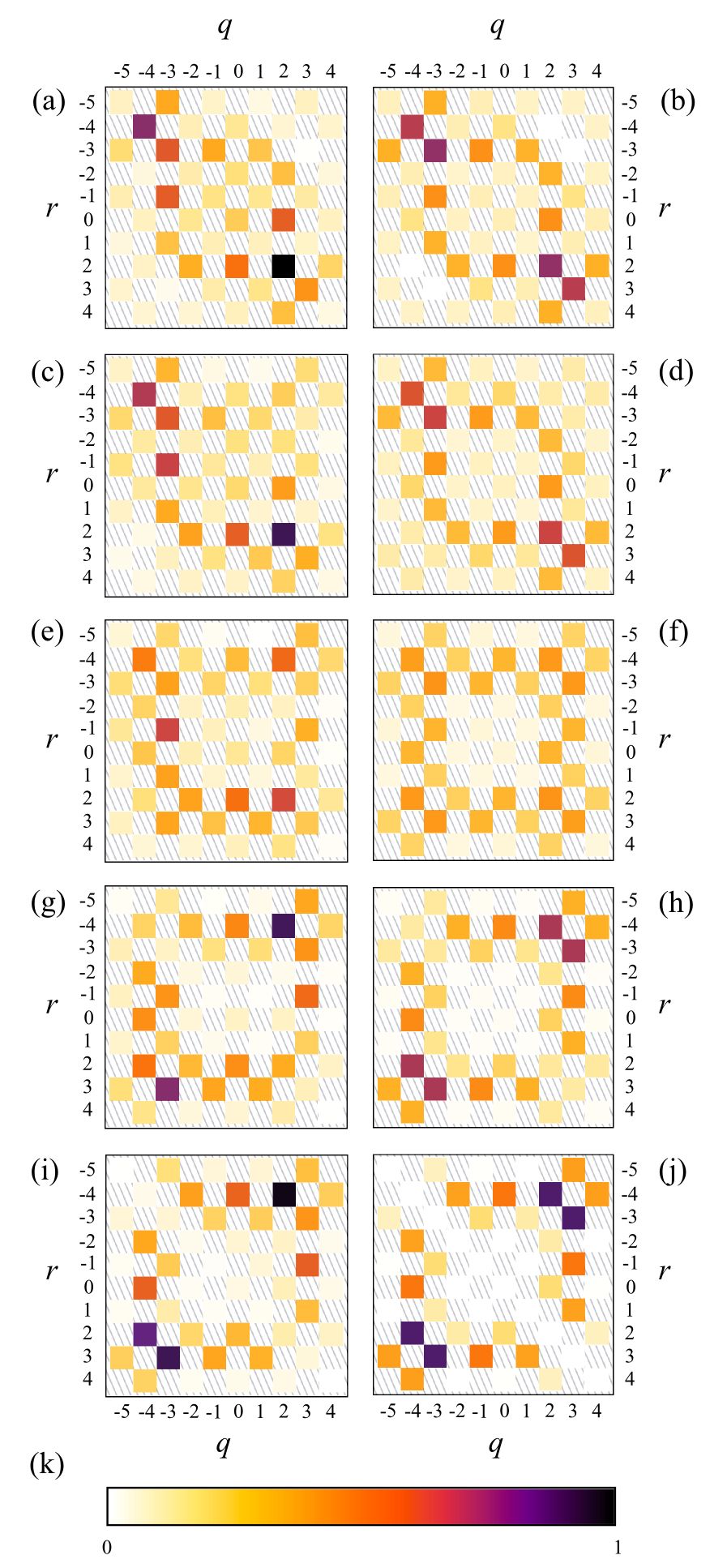}
\caption[]{\footnotesize{Experimental simulation of particle statistics correlation functions for input ($-1$, $0$). Experimental correlation matrices (left: a,c,e,g,i) for simulating quantum interference of non-interacting particle pairs with five different particle statistics. Ideal correlation matrices for this experiment are plotted (right: b,d,f,h,j) using Eq.~(\ref{2_part_correlation_fn}). The data ranges from corresponding to Bose-Einstein statistics ($\phi = 0$, (a,b)) through fractional statistics ($\phi = \pi/4$, (c,d); $\phi = \pi/2$, (e,f); $\phi = 3\pi/4$, (g,h)) to Fermi-Dirac statistics ($\phi = \pi$, (i,j)). (k) The correlation matrices are displayed as normalised probability distributions with white for probability $P^\phi_{r,q} = 0$ and black for $P^\phi_{r,q} = 1$. Solid coloured squares represent measured correlations, hatched squares correspond to coincidences that cannot be measured due to mis-match between fibre array separation and waveguide pitch (see Supplementary Information). Coincidence rates are corrected for relative detection efficiency.}}
\label{fig_data110}
\end{figure}

\begin{figure}[htp!]
    \centering
    \includegraphics[width = 8cm]{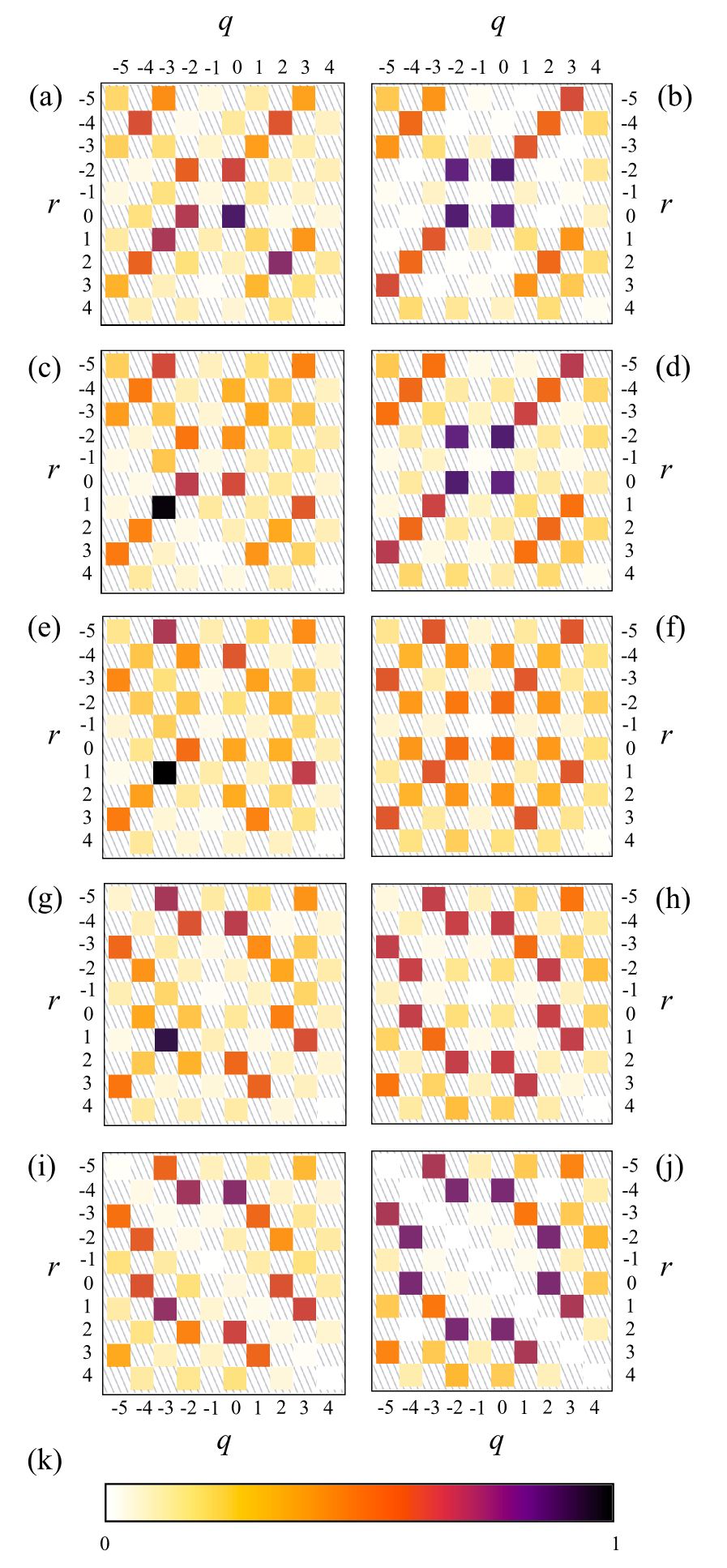}
\caption[]{\footnotesize{
Experimental simulation of particle statistics correlation functions for input ($-1$, $1$) with data laid out as for Fig.~\ref{fig_data110}.}}
\label{fig_data101}
\end{figure}

We measured two sets of correlation matrices for two different inputs: immediate neighbouring waveguides  in the centre of the array $(j,k)=(-1,0)$ (Fig.~\ref{fig_data110}); and next but one neighbouring waveguides $(j,k)=(-1,1)$, with vacuum input in waveguide $0$ (Fig.~\ref{fig_data101}). Each $P^\phi_{r,q}$ entry is the probability to detect a horizontal and vertical polarised photon at outputs $r$ and $q$, respectively. Five cases are shown for $\phi = 0, \pi/4, \pi/2, 3\pi/4, \pi$ (left: (a), (c), (e), (g), (i)). The corresponding ideal correlation matrices given by $\Gamma_{r,q}^\phi$ in Eq.~(\ref{2_part_correlation_fn}) are also plotted (right: (b), (d), (f), (h), (j)). 
For the $(j,k)=(-1,0)$ input (Fig.~\ref{fig_data110}) we observe the expected bunching behaviour of bosons for $\phi = 0$ which is contrasted with the observation of the Pauli-exclusion in the diagonal elements of the matrices for $\phi = \pi$. We note that a form of exclusion on the diagonal entries $P^\phi_{r,r}$ would also occur for experiments with hard core bosons, but the remaining correlation matrix would still be dependent upon Bose-Einstein exchange statistics. The spectrum of phase $0<\phi<\pi$ exhibits a variable Pauli-exclusion and a continuous transition between simulation of Bose-Einstein and Fermi-Dirac quantum interference.
In the bosonic case, we observe a less pronounced bunching behaviour for the $(j,k)=(-1,1)$ input (Fig.~\ref{fig_data101} (a)) than for the $(j,k)=(-1,0)$ input (Fig.~\ref{fig_data110} (a)), as expected \cite{br-prl-102-253904}. This behaviour is inverted in the fermionic case, since the particels tend to propagate to the same peak of the walk (while still not occupying the same waveguide) for the $(j,k)=(-1,1)$ input (Fig.~\ref{fig_data101} (i)), and tend to propagate in different peaks for the $(j,k)=(-1,0)$ input  (Fig.~\ref{fig_data110} (i)). This is the result of the different phases in the Hamiltonian that describes the quantum walk evolution.

Agreement between the ideal implementation of our scheme, with perfect realisations of $A^a$ and $A^b$, and the data in Figs.~\ref{fig_data110},\ref{fig_data101} is quantified in Table~\ref{similarity_tab} using the similarity $S$ between the ideal correlation matrix $\Gamma^\phi$ and the measured distribution of detection statistics $P^\phi$: $S= (\sum _{r,q}\sqrt{\Gamma^\phi_{r,q} P^\phi _{r,q}})^2/{\sum _{r,q}\Gamma^\phi _{r,q} \sum _{r,q}P^\phi _{r,q}}$, a generalisation of the average (classical) fidelity between two probability distributions that ranges from $0\%$ (anti-correlated distributions) to $100\%$ for complete agreement. We also calculated $S$ when we deliberately made the photons distinguishable by temporally delaying one with respect to the other. This enabled  physical simulation of the classical correlations $\Gamma^C = |A_{r,j}A_{q,k}|^2+|A_{r,k}A_{q,j}|^2$ that are independent of particle statistics. For these measurements, we found agreement with the ideal model of $S = 97.4\pm 0.7\%$ and $S = 97.5\pm 0.6\%$ for inputs $(-1,0)$ and $(-1,1)$ respectively. These data indicate that $A^a$ and $A^b$ are near identical while keeping crosstalk between horizontal and vertical polarisation to a minimum. While non-unit similarity ($S = 100\%$) in Table~\ref{similarity_tab} is attributed to non-perfect entanglement generation, the high level of agreement between the ideal and measured cases indicates accurate quantum simulation of arbitrary quantum interference in a uniform coupled oscillator Hamiltonian.


\begin{table}[p!]
\begin{center} 
\begin{tabular}{ | l |  c | c | }
\hline Phase, $\phi$ &  S (\%) for input $(-1,0)$ & S (\%) for input $(-1,1)$ \\
\hline $0$ & $95.8 \pm 0.6$ & $90.9 \pm 0.5$ \\
\hline $\pi/4$ & $95.2 \pm 0.7$ & $93.6 \pm 0.6$ \\
\hline $\pi/2$ &$94.1 \pm 0.7$ &$93.5 \pm 0.6$ \\
\hline $3\pi/4$ & $94.4 \pm 0.7$ & $94.1 \pm 0.6$ \\
\hline $\pi$ & $91.6 \pm 0.7$ &  $92.7 \pm 0.6$\\
\hline
\end{tabular} 
\end{center}
\caption[]{\footnotesize{{Similarity $S$ for each experimental simulation of particle quantum interference with parameter $\phi$, for input $(-1,0)$ given in Fig.~\ref{fig_data110} and input $(-1,1)$ given in Fig.~\ref{fig_data101}.}}}
\label{similarity_tab}
\end{table}

\subsection*{Generalization to N particles}

Using multi-partite entanglement allows simulation of quantum interference of an arbitrary number $N$ of identical non-interacting particles. 
$N$ indistinguishable particles with exchange statistics $\phi$, subjected to mode transformation $A$ at inputs $\vec{\nu}=\{\nu_1,\nu_2,...,\nu_N\}$, are correlated due to quantum interference across outputs $\vec{\mu}=\{\mu_1,\mu_2,...,\mu_N\}$ according to
\begin{eqnarray}
\Gamma^\phi_{\vec{\mu}} = \Big| \sum\limits_{\sigma_\nu \in S_N} 
e^{i\tau\left(\sigma_\nu\right)\phi}
\prod\limits_{j=1}^N A_{\mu_j, \sigma_{\nu_j}} \Big|^2,
\label{n_particlestatistics}
\end{eqnarray}
where $\sigma_\nu$ is an element of the permutation group $S_N$ acting on $\vec{\nu}$ and $\tau\left(\sigma_\nu\right)$ is the minimum number of neighbouring swaps that maps $\sigma_\nu$ to $\vec{\nu}$; ${\sigma_\nu}_{j}$ denotes the element $j$ of ${\sigma_\nu}$. For $\phi = 0$ and $\phi = \pi$, Eq.~(\ref{n_particlestatistics}) is the modulus-square of the permanent and determinant respectively of an $N\times N$ subset of $A$. For $0\!<\!\phi\!<\!\pi$ Eq.~(\ref{n_particlestatistics})---which is the modulus-square of the imminent of the $N\times N$ subset of $A$---is derived in the Supplementary Information.
The multi-particle correlation function Eq.~(\ref{n_particlestatistics}) is simulated by subjecting the generalised $N$-partite, $N$-level entangled state 
\begin{eqnarray}
\frac{1}{\sqrt{N!}}\sum\limits_{{\sigma_\nu}\in S_{N}}  e^{i\tau\left(\sigma_\nu\right)\phi} \prod\limits_{j=1}^{N} {a_{{\sigma_\nu}_{j}}^{(j)^\dagger}}\ket{0}, 
\label{general_singlet}
\label{big_entang_state}
\end{eqnarray}
to $N$ copies of $A$ such that one particle ${a^{(j)}}^\dagger$ is launched into each copy $A^{(j)}$.
For example, for inputs $j,k,l$ and $N=3$, the required entangled input state is equivalent to three entangled qutrits, shared across three copies of $A$ denoted $A^a,A^b$ and $A^c$:
\begin{eqnarray}
&&\frac{1}{\sqrt{6}}\Big(a^\dagger_j b^\dagger_k c^\dagger_l + e^{i\phi} a^\dagger_j b^\dagger_l c^\dagger_k + e^{i\phi} a^\dagger_k b^\dagger_j c^\dagger_l \nonumber \\
 &&+ e^{i 2\phi} a^\dagger_k b^\dagger_l c^\dagger_j + e^ {i 2\phi}a^\dagger_l b^\dagger_j c^\dagger_k + e^{i3\phi} a^\dagger_l b^\dagger_k c^\dagger_j\Big)\ket{0}
\end{eqnarray}
$N$-fold coincidence detection across outputs $\mu_j$ of $A^{(j)}$ equals $\frac{1}{N!} \Gamma^\phi_{\vec{\mu}}$. Note as for the case with two particles, the $1/N!$ pre-factor can be canceled by considering all $N!$ permutations of the outputs, 
therefore directly simulating quantum interference of $N$ bosons ($\phi = 0$) or $N$ fermions ($\phi = \pi$). For example, in the 2-particle case, the $1/2$ pre-factor in Eq.~(\ref{emulationeqn}) is canceled by summing both the detection outputs of a particle at $r$ on $A^a$, $q$ on $A^b$ and at $r$ on $A^b$, $q$ on $A^a$. All other values of $\phi$ allow continuous transition between Fermi and Bose quantum interference. Note that the approach of initialising the simulation with a pre-symmetrized (or pre-antisymmetrized) state is in the spirit of first quantised quantum simulation \cite{ka-arc-62-185}, for which quantum algorithms have been designed to efficiently generate the required states \cite{ab-prl-79-2586,wa-jcp-130-194105} (see Supplementary Information). In a photonic realization of the proposed sheme, the required N-particle entangled state can be generated using additional photons and feedforward to implement quantum gates following the KLM scheme\cite{kn-nat-409-46}.

\section*{Discussion}

Correlations resulting from quantum interference of multiple indistinguishable non-interacting bosons in large linear optical circuits cannot be efficiently simulated with classical computers \cite{scheel-permanents, aaroson-acm, aa-arXiv-1011.3245}. Large scale repeated quantum interference is achievable with photons in waveguides \cite{po-sci-320-646,pe-sci-329-1500,ma-prl-107-163602,pe-ncomm-2-224}, which is likely to provide the platform for the realisation of such a boson scattering device, to realise an
engineered system that can outperform the limits of classical computation.
A major issue will be verifying boson scattering correlations that exceed the capabilities of conventional computers. In contrast, the analogous situation for non-interacting fermions is efficient with classical computers \footnote{Currently it is not as straightforward to perform quantum interference of fermions \cite{li-nat-391-263,ro-nat-444-733} as it is with photons in integrated optics.}.
The ability to tune continuously between tractable and intractable phenomena with a single setup could be essential for verification of bosonic scattering. 

Other possible applications for simulating 
particle statistics with photonics include preparing appropriately symmetrized states in quantum chemistry simulations \cite{ka-arc-62-185} and combinatorics, where exchange statistics have been shown to alter mathematical graph structure \cite{ha-procrsoca-advonline}---three-dimensional waveguide architectures \cite{ke-pra-81-023834} may be important here.
Our simulation of arbitrary fractional exchange statistics also provides a novel method to study fractional quantum interference phenomena \cite{ke-ncomm-2-361}; whether our approach can be used in the context of topological fault tolerant quantum computation with photons \cite{lu-prl-102-030502,pa-njp-11-083010} is an open question.

The method we report can be directly applied to any quantum process realisable with linear optics \cite{re-prl-73-58} and is not reliant on using polarisation---the entangled state could be encoded in time-bin or interferometrically stable path with reconfigurable waveguide circuits \cite{ma-natphot-3-346}. 
In order to introduce interactions between simulated particles, it may be possible to use effective non-linearities using auxiliary photons \cite{kn-nat-409-46,ok-pnas-108-100067}. In the long term, the method we have demonstrated here could be applied to other quantum platforms (consisting of fermions or bosons) where nonlinear behaviour may be more readily exploited.

\section*{Acknowledgements}
This work was supported by EPSRC, ERC, US AFOSR, NSQI and European projects PHORBITECH,  QUANTIP. J.C.F.M. is supported by a Leverhulme Trust Early Career Fellowship. A. Pe. holds a Royal Academy of Engineering Research Fellowship. J.L.OB. acknowledges a Royal Society Wolfson Merit Award and a Royal Academy of Engineering Chair in Emerging Technologies.

\section*{Methods}

The photon pairs are generated in a 2mm thick, nonlinear bismuth borate $\textrm{BiB}_3\textrm{O}_6$ (BiBO) crystal phase matched for non-colinear type-I spontaneous parametric downconversion, pumped by a 60 mW 404 nm continuous wave diode laser. The photons are spectrally filtered using 2nm full width at half maximum interference filters centred on 808nm and collected with aspheric lenses focusing on polarisation maintaining fibres (PMFs). The typical two-photon count rate directly from the source with two $~60\%$ efficient single photon counting modules (SPCMs) is 50kHz. For a coherent input state, arrival time of the two photons at the input of the waveguide array is made identical via characterisation by performing a Hong-Ou-Mandel experiment \cite{appho-prl-59-2044}. The polarisation state $\ket{D} = (\ket{H} + \ket{V})/\sqrt{2}$ is then prepared for both photons using a half wave-plate with optic axis set to $\pi/8$ radians. The polarisation in the left arm is matched and purified to that of the right using a rotated polarisation beam splitter in transmission. The two photon state launched into the waveguide array is then $(\ket{H}_j + \ket{V}_j)(\ket{H}_k + e^{i(\phi+\epsilon)}  \ket{V}_k)/2$. The collecting PMF and the birefringence of the waveguide array preserve horizontal and vertical polarisation components for both photons and prohibit crosstalk between horizontal and vertical light; PMF-coupled polarisation beam splitters at the output of the waveguide array allow two-photon correlated detections across the two orthogonal polarisations to project the initial input state to be the entangled state $(\ket{H}_j\ket{V}_k + e^{i(\phi+\epsilon)} \ket{V}_j\ket{H}_k)/\sqrt{2}$; the extra phase $\epsilon$ is compensated using a $Z(\theta)$ rotation on the right arm of the photon source comprised of two quarter waveplates (optic axis at $\pi/4$ radians) with a half waveplate between (optic axis at $-(\pi+\phi+\epsilon)/4$ radians, see Fig.~1 of the main text).
\newline

The waveguide array was fabricated from silicon oxynitride  (SiON) thin films were grown by plasma enhanced chemical vapour deposition (PECVD) onto thermally oxidised $<\!\!100\!\!>$ oriented 100-mm silicon wafers applying an Oxford 80 Plasmalab Systems. A deposition process based on 2\% $\textrm{SiH}_4/\textrm{N}_2$ and $\textrm{N}_2\textrm{O}$ was performed at 60W power, 187.5 kHz frequency, 650 mTorr pressure and $300^\circ$C substrate temperature. The SiON films were annealed for 3hours at $1150^\circ$C in a nitrogen atmosphere. By optical inspection at 632 nm wavelength a film thickness of $d = 603$ $(\pm3)$ nm and a refractive index (for TE) of $n_{\textrm{\tiny{TE}}} = 1.5232$ $(\pm1\times10^{-3})$ were measured. The channel waveguides were obtained from standard lithography followed by reactive ion etching in $\textrm{CHF}_3 / \textrm{O}_2$ chemistry at 350 W plasma power. The etch depth $(d=613\pm 5\textrm{nm})$ was confirmed by profilometry. A 5-$\mu$m thick upper cladding layer was deposited in two steps applying a Low Pressure CVD TEOS-process and standard PECVD $\textrm{SiO}_2$ deposition. The upper cladding was annealed at $1150^\circ$C for 3 hours. For a core width of 1.8-$\mu$m the total birefringence of the waveguide, including material and form birefringence \cite{app-wo-JLT-23}, is of the order $\Delta n=n_{\textrm{\tiny{TM}}}-n_{\textrm{\tiny{TE}}}=2*10^{-4}$. This value provides polarization preserving propagation for the TE and TM photons throughout the device. Moreover, the birefringence is small enough to ensure a nearly identical mode overlap, and coupling ratio, between different waveguides in the coupled array for TM and TE modes. This implies that the processes $A^a = A^{TE}$ and $A^b=A^{TM}$ are matched, as required and verified in the reported experiment. The coupled waveguide arrays were accessed with lithographically patterned bends (minimum bend radius of 600$\mu$m) that gradually mapped the 250$\mu$m pitched input (for 250$\mu$m pitched PMF array access) to the 2.8$\mu$m pitch in the evanescent coupling region. The coupling region is 350$\mu$m long, before bending gradually to an output pitch of 125$\mu$m which was accessed by 250$\mu$m pitched arrays of PMF; the output measurements were alternated between access to all odd labeled waveguides and all even labeled waveguides. The optical chips were diced from the wafer.
\newline

\bibliographystyle{unsrtnat}

\newpage

\section*{Supplementary Information}

\noindent{\textbf{Operator for continuous, constant nearest neighbour coupling.}} Evanescently coupled uniform waveguide arrays are modelled with the coupled oscillator Hamiltonian 
\begin{eqnarray}
H = \sum_j \beta a^\dagger_j a_j + C  a^\dagger_{j-1} a_j  + C a^\dagger_{j+1} a_j
\label{app_ham}
\end{eqnarray}
for waveguide propagation constant (site potential) $\beta$ and constant coupling coefficient $C$, where $\beta,C\in \mathbb{R}$. In the basis of single position (or mode) states $\{\ket{1}_1,\ket{1}_2,...\ket{1}_n\}$, the Hamiltonian is represented by a tri-diagonal $M\times M$ matrix 
\begin{eqnarray}
H_{i,j} = \left\{\begin{array}{l l} \beta, & i = j\\ C, & i = j\pm 1 \end{array}\right. .
\end{eqnarray}
Unitary evolution of this Hamiltonian after time $T=cz$ is given by $U= e^{i H T}$, where $z$ is propagation distance in the waveguide and $c$ is the effective propagation speed which is equivalent to a continuous time quantum walk on a one dimensional graph \cite{apppe-prl-100-170506}. 

 The device reported is a uniform array of $M=21$ waveguides. The propagation length $z$ is such that the linear ballistic propagation resulting from launching light into the central waveguides \cite{apppe-prl-100-170506} does not extend beyond ten central waveguides which are accessed in the experiment. For computing similarity between experiment and theory the matrix $A$
is modelled by extracting the central $10\times 10$ submatrix of a $21\times 21$ unitary matrix $U = e^{iT H}$, computed using Eq.~\ref{app_ham} with parameters $T=13.9$, $\beta = 0$ and $C = 0.15$; the theoretical model assumes the array to be uniform and to have zero loss. Both assumptions can give rise to $S<100\%$ in our analysis.
\newline

\noindent\textbf{2-particle correlation functions.}
Quantum interference between two indistinguishable particles---modelled by creation operators $a^\dagger_j$ and annihilation operators $a_j$ acting on labeled modes $j$ in a finite set of modes associated to system $a$---is quantified by the correlation function $\Gamma_{r,q}^\phi$ and has been derived for photons (bosons) \cite{app-ma-apb-60-s111} and fermions \cite{app-li-njp-7-155} in multi-port networks. Here we review derivation and adapt for fractional statistics that allow continuous deformation between Bose-Einstein and Fermi-Dirac statistics. 

Consider two indistinguishable particles that are governed by the relations
\begin{eqnarray}
&& a^\dagger_j a^\dagger_k - e^{i \phi} a^\dagger_k a^\dagger_j = 0 \label{com1}\\ 
&& a_j a_k - e^{i \phi} a_k a_j = 0\label{com2}\\
&& a_j a^\dagger_k - e^{i \phi} a^\dagger_k a_j = \delta_{j,k}\label{com3}
\end{eqnarray}
for Bose-Einstein ($\phi = 0$) and Fermi-Dirac ($\phi = \pi$) statistics, governing exchange according to the permutation group \cite{fey-stat-mech}. We can also define particles that undergo a fractional phase change $0<\phi<\pi$ with $j<k$, on exchanging two such particles according to braiding operations defined as `clockwise'. For $j>k$, anti-clockwise braids are defined with $-\pi<\phi<0$.

A mode transformation $A$ on a state of two indistinguishable particles created in modes $j$ and $k$, maps the initial state $a^\dagger_j a^\dagger_k\ket{0}$ to a linear superposition of two particle states according to
\begin{eqnarray}
a^\dagger_j a^\dagger_k\ket{0} \stackrel{A}{\rightarrow}\sum_{s,t}A_{s,j}A_{t,k}a^\dagger_s a^\dagger_t\ket{0},
\label{AppUnitaryTransform}
\end{eqnarray}
where $A$ is expressed as a matrix in the mode basis. Correlated detection at $r$ and $q$ will occur as a result of two possible events: (i.) the particle at $j$ is mapped to position $r$ and the particle at $k$ is mapped to $q$ with amplitude $A_{r,j}A_{q,k}$; (ii.) the particle at $j$ is mapped to position $q$ and the particle at $k$ is mapped to $r$ with amplitude $A_{q,j}A_{r,k}$. Labelling modes such that $j<k$ and $r<q$ implies (ii.) is equivalent to at least one exchange of the position of the two particles. Here we define (i.) corresponds to no exchange while (ii.) corresponds to a single swap or braid operation according to (\ref{com1}-\ref{com3}).

Since the two particles are indistinguishable, quantum mechanics dictates outcomes (i.) and (ii.) are indistinguishable and hence the complex probability amplitudes $A_{r,j}A_{q,k}$ and $A_{q,j}A_{r,k}$ interfere in the correlation across $r$ and $q$. The correlated detection outcome $\Gamma^\phi_{r,q}$ is therefore computed by projecting the state $\bra{0}a_r a_q$ on the output state (\ref{AppUnitaryTransform}), using the inner product 
\begin{eqnarray}
\bra{0} a_r a_q a^\dagger_j a^\dagger_k \ket{0} = \delta_{q,j}\delta_{r,k} + e^{i \phi }\delta_{r,j}\delta_{q,k},
\label{AppFermiStat}
\end{eqnarray}
which follows from the relations (\ref{com1}-\ref{com3}). Combining expressions (\ref{AppUnitaryTransform}) and (\ref{AppFermiStat}) therefore yields
\begin{eqnarray}
\Gamma^\phi_{r,q} =  \left| A_{r,j} A_{q,k} +e^{i\phi} A_{q,j} A_{r,k} \right|^2
\label{arbstats}
\end{eqnarray}
for exchange statistics parameterised by $\phi$. Note that for fermions ($\phi = \pi$) the diagonal elements of the matrix $\Gamma^\pi_{r,r} = 0$ agree with the Pauli-exclusion principle. 
\newline

\noindent\textbf{$N$-particle correlation functions.} Consider $N$ particles with exchange parameter $\phi$ and governed by the relations (\ref{com1}-\ref{com3}) which are launched into inputs labelled by tuple $\vec{\nu}=\{\nu_1,\nu_2,...,\nu_N\}$ of mode transformation $A$. The $N$-fold correlation function for detection across output $N$-tuple $\vec{\mu}=\{\mu_1,\mu_2,...,\mu_N\}$ is derived for multi-particle quantum walks with both Fermi-Dirac and Bose-Einstein statistics \cite{appma-arXiv-1009-5241}. 

The input state is mapped by arbitrary $A$ to a linear superposition of $N$-particle states at the output of $A$ according to
\begin{eqnarray}
\prod_{j=1}^N a^\dagger_{\nu_j}\ket{0} \stackrel{A}{\rightarrow} \prod_{j=1}^N \left(\sum_{p=1}^{N} A_{p,\nu_{j}}a^\dagger_{p}\right) \ket{0}.
\label{npart_map}
\end{eqnarray}
Defining $\nu_1\!\!<\!\!\nu_2\!\!<\!\!...\!\!<\!\!\nu_N$ and $\mu_1\!\!<\!\!\mu_2\!\!<\!\!...\!\!<\!\!\mu_N$ we can consider the transformation $A$ maps the $N$ particles at $\vec{\nu}$ to the output $\vec{\mu}$ in $N!$ indistinguishable possibilities. One of these possibilities maps $\nu_j$ to $\mu_j$ for all $j$ which we define to not exchange a single pair of particles. The remaining $N!-1$ map  $\nu_j$ to $\sigma_{\mu_j}$ for all $j$ for some permutation $\sigma_{\mu}$ of $\mu$ and are defined to exchange particles with the minimum number of pairwise transpositions in clockwise braids that transform $\nu$ to $\sigma_{\mu}$. Each braid between any two particles makes the combined wave-function acquire a phase $\phi$; $m$ clockwise braids that are the minimum number required to map $\nu$ to $\sigma_{\nu}$ therefore cause the global wave-function to acquire phase $m\phi$. From this definition, we obtain an $N$-particle generalisation of expression (\ref{AppFermiStat}):
\begin{eqnarray}
\Big(\!\!\bra{0} \prod_{s=1}^N a_{\mu_s}\Big) \Big(\!\prod_{j=1}^N {a^\dagger}_{\nu_j}\ket{0}\! \! \Big)\! =\!\!\!\!\sum_{\sigma_{\nu}\in S_n} \!e^{i\tau(\sigma_\nu)\phi}\prod^N_{j=1}\delta_{{\mu_j},{\sigma_\mu}_j}\!,
\label{AppFermiStatgen}
\end{eqnarray}
where $\tau(\sigma_\nu)$ is the minimum number of pairwise transpositions (neighbouring swaps) relating $\sigma_\nu$ to $\vec{\nu}$. Projection onto the output state (\ref{npart_map}) with $N$-particles at $\vec{\mu}$ uses relations (\ref{com1}-\ref{com3}) and (\ref{AppFermiStatgen}) for a fractional parameter ($0<\phi<\pi$) to yield
\begin{eqnarray}
\Gamma^\phi_{\vec{\mu}} = \left| \sum\limits_{\sigma_\nu \in S_n} e^{i\tau\left(\sigma_\nu\right)\phi}
\prod\limits_{j=1}^n A_{\mu_j, \sigma_{\nu_j}} \right|^2.
\label{appn_particlestatistics}
\end{eqnarray}
Varying $\phi$ allows continuous deformation between Bose-Einstein ($\phi= 0$) and Fermi-Dirac ($\phi = \pi$) quantum interference.
\newline

\noindent\textbf{Generating $N-$partite, $N-$level entanglement.} A quantum algorithm for anti-symmetrising arbitrary products of $n$ wave functions $\ket{\rho_1,\rho_2,...,\rho_n}_A$ is presented \cite{app-ab-prl-79-2586} that would use three registers of $n$, $m-$digit quantum words or ``qu-words''   (strings of $n \log_2 m$  qubits). This was then simplified \cite{app-wa-jcp-130-194105} to use two registers of qubits $A$ and $B$; the first encodes the wavefunctions $\ket{\rho_1,\rho_2,...,\rho_n}$. 
The second register is mapped to a symmetrized state with $O(n(\ln m)^2)$ local operations followed by $O(n^2\ln m)$ controlled operations leaving the second register in the state
\begin{eqnarray}
\frac{1}{\sqrt{N!}}\sum_{\sigma\in S_N} \ket{\sigma(1,...,N)}_{B}
\label{app_eq_sym_state}
\end{eqnarray}
where $\sigma(1,..,N)$ is a permutation $\sigma \in S_N$ of the $N-$tuple $\{1,2,...,N\}$. The state Eq.~\ref{app_eq_sym_state} is then antisymmetrised by a sorting algorithm using $O(n \ln n)$ operations, applying the same swaps (via controlled operations) and phase shifts to the first register to antisymmetrize $\ket{\rho_1,\rho_2,...,\rho_n}_A$. For our purpose, we require only the one register $B$, since the state we wish to generate is known and of the form
\begin{eqnarray}
\ket{\psi_N \left(\phi \right)} & = & \frac{1}{\sqrt{N!}}\sum\limits_{{\sigma_\nu}\in S_{N}}  e^{i\tau\left(\sigma_\nu\right)\phi} \prod\limits_{j=1}^{N} {a_{{\sigma_\nu}_{j}}^{(j)^\dagger}}\ket{0}\nonumber\\
& = & \frac{1}{\sqrt{N!}}\sum\limits_{{\sigma_\nu}\in S_{N}}  e^{i\tau\left(\sigma_\nu\right)\phi} \ket{\sigma_\nu(1,2,...,N)}
\label{general_singlet}
\label{big_entang_state}
\end{eqnarray}
where $\sigma(1,..,N)$ can be decomposed into $\tau(\sigma)$ adjacent transpositions. 

\begin{figure}
    \centering
    \includegraphics[width = 8cm]{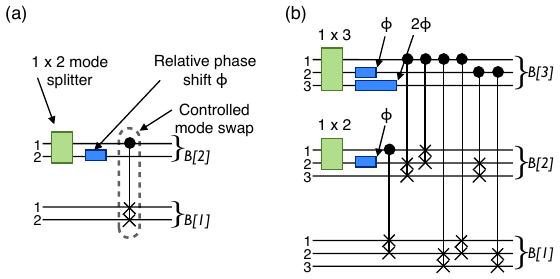}
\caption[]{\footnotesize{Qudit circuits for geneating $\ket{\psi_N(\phi)}$ for (a) $N=2$ and (b) $N=3$.
Qudits are initialised in the state $\ket{1}_{B[N]}\otimes ... \otimes\ket{1}_{B[1]}$, where $B[i]$ denotes the $i^\textrm{th}$ qudit in the register. The green squares represent $1\times k$ mode splitters that can be decomposed into $k-1$ two-mode splitters. Blue rectangles represent a relative phase shift with labelled phase $j\phi$. Each controlled swap between two modes denoted `$\times$' are conditional on an excitation in the corresponding mode denoted denoted by a `$\bullet$'.}}
\label{fig_stategenerator}
\end{figure}

The approach shown in Fig.~\ref{fig_stategenerator} represents each $i^{th}$ qu-word (denoted $B[i]$) in the register  as an $N-$level qudit,
for the examples $N=2,3$, and generates the state $\ket{\psi_N \left(\phi \right)}$ in an iterative process, using $\sum_{i=1}^{N-1} i \sum_{j=1}^{i} j < N^3$ controlled operations.
Here each controlled operation is counted as one swap operation of two modes in a given qudit, conditional on an excitation in one of the $N$ levels of the control qudit. Realisation with linear optics for example, could use a heralded KLM CNOT gate \cite{app-ra-pra-65-012314,app-ok-arxiv}, each requiring two extra (ancilla) photons, and an interferometric structure of seven beamsplitters. 

As with Refs.~\onlinecite{app-ab-prl-79-2586,app-wa-jcp-130-194105}, the $N$ qudits are initialised in superposition using local operations (mode splitters; $MS$) 
\begin{eqnarray}
&&\ket{1}_{B[N]}\otimes ... \otimes\ket{1}_{B[1]}\\
&&\stackrel{MS}{\longrightarrow}\!\frac{1}{\sqrt{N!}}\!\Bigg(\!\sum_{k=1}^{N}\ket{k}_{B[N]}\!\!\Bigg)\!\!\otimes\!\!\Bigg(\!\sum_{k=1}^{N-1}\ket{k}_{B[N-1]}\!\Bigg)\!\!\otimes \!... \!\otimes\!\ket{1}_{B[1]}\nonumber
\end{eqnarray}
Note that this step puts the state in the correct number of $N!$ terms required for the eventual entangled state. Using linear optics, this would use one photon for each qudit and $\sum_{j=1}^N (j-1) = O(N^2)$ two-mode beam splitters (with appropriate reflectivities). 

The next step in generating $\ket{\psi_N(\phi)}$ is to 
introduce the phase relationship $\phi$ according to
\begin{eqnarray}
\frac{1}{\sqrt{N!}}\!\Bigg(\!\sum_{k=1}^{N}e^{i(k-1)\phi}\ket{k}_{B[N]}\!\!\Bigg)\!\!\otimes\!\!\Bigg(\!\sum_{k=1}^{N-1}\!\!\!\!\!\!\!&&e^{i(k-1)\phi}\ket{k}_{B[N-1]}\!\!\Bigg)\!\!\otimes \!...\!\nonumber\\
&&...\! \otimes \!\ket{1}\!_{B[1]}\\
&&\nonumber
\end{eqnarray}

\noindent which uses $\sum_{j=1}^N (j-1)= O(N^2)$ local phase shift operations. A set of $\sum_{i=1}^{N-1} i \sum_{j=1}^{i} j = O(N^4)$  controlled operations are then applied iteratively to generate $\ket{\psi_N(\phi)}$; this process can be thought of as performing a series of controlled operations on the intermediate states
\begin{eqnarray}
\frac{1}{\sqrt{q+1}}\sum_{j=1}^{q+1}e^{i(j-1)\phi}\ket{j}_{B[q+1]} \otimes\ket{\psi_q (\phi)}_{B[q,...,1]}
\end{eqnarray}
for each step of the iteration $q=2,...,N-1$. This is verified for $q=2$ (see Fig.~\ref{fig_stategenerator}(a) and Fig.~\ref{fig_stategenerator}(b)). Next we assume inductively the process works for some $q=r$ to generate $\ket{\psi_q(\phi)}$. For $q=r+1$, the process would begin with the state
\begin{eqnarray}
&&\frac{1}{\sqrt{r\!+\!1}}\sum_{j=1}^{r+1}\!e^{i(j-1)\phi}\!\ket{j }_{B[r+1]}\! \otimes\!\ket{\psi_r (\phi)}_{B[r,..,1]}\\
\!\!&&=\!\! \frac{1}{\sqrt{r\!+\!1}}\!\!\sum_{j=1}^{r+1}\!e^{i(j-1)\phi}\!\ket{j}_{B[r+1]}\! \otimes\! \frac{1}{\sqrt{r!}}\! \!\sum_{\sigma \in S_r} \!\ket{\sigma(1,\!..,r\!)}_{B[r,..,1]}\nonumber
\end{eqnarray}
A series of controlled qudit shift operations are then applied, such that an excitation of mode $j$ of the $B[r+1]$ qudit shifts each of the labeled modes $\{j,...,r\}$ to $\{j+1,...,r+1\}$ in the target qudits labelled $\{B[r],\ldots B[1]\}$, such that the overall state evolves to
\begin{widetext}
\begin{eqnarray}
\frac{1}{\sqrt{(r+1)!}}\sum_{\sigma \in S_r} e^{i\tau_\nu(\sigma)\phi} \Bigg\{ \ket{1}_{B[r+1]} \ket{\sigma(2,...,r+1)}_{B[r,...,1]} + e^{i\phi}\ket{2}_{B[r+1]} \ket{\sigma(1,3,4,...,r+1)}_{B[r,...,1]}\nonumber\\
\ldots+e^{i(j-1)\phi}\ket{j}_{B[r+1]} \ket{\sigma(1,2,3,...,j-1,j+1,...,r+1)}_{B[r,...,1]} +\ldots\nonumber\\
\ldots+e^{ir\phi}\ket{r+1}_{B[r+1]} \ket{\sigma(1,2,...,r)}_{B[r,...,1]} \Bigg\}
\label{inductive_step}
\end{eqnarray}
\end{widetext}

The tuple $\{j,1,2,...,j-1,j+1,...,r+1\}$ is $(j-1)$ adjacent transpositions from the tuple $\{1,2,...,r+1\}$. Since the permutation $\sigma(1,2,..,j-1,j+1,...,r+1)$ is $\chi(\sigma)$ adjacent transpositions away from the ordered set $\{1,2,...,j-1,j+1,...,r+1\}$, it follows that the tuple $\{j,\sigma(1,2,...,j-1,j+1,...,r+1)\}$ is $(j-1 + \chi(\sigma))$ adjacent transpositions different from $\{1,2,..,r+1\}$. Hence the Eq.~\ref{inductive_step} is equivalent to the state
\begin{eqnarray}
\ket{\psi_{r+1} (\phi)} \! = \! \frac{1}{\sqrt{(r\!+\!1)!}} \! \sum_{\gamma \in S_r} \! e^{i\tau_\nu(\gamma)\phi} \! \ket{\gamma(1,..,r\!+\!1)}_{B[r\!+\!1,..,1]}\nonumber\\
\end{eqnarray}
as required. Therefore, by induction we have an iterative process that generates $\ket{\psi_N(\phi)}$ for arbitrary size $q=N$, requiring $O(N^2)$ local operations, and $O(N^4)$ controlled swap operations.

\end{document}